\documentclass[aps,prb,twocolumn,amsfonts,a4paper,floatfix,nofootinbib]{revtex4}

\usepackage{physics}
\usepackage{mathtools}
\usepackage{graphicx}

\begin{document}

\title[Applicability and limitations  of CPT]{Applicability and limitations  of cluster perturbation theory for Hubbard models}

\author{{Nicklas} {Enenkel}}
\affiliation{HQS Quantum Simulations GmbH, Rintheimer Stra{\ss}e 23, 76131 Karlsruhe, Germany}
\affiliation{Institut für Theoretische Festkörperphysik, Karlsruhe Institute of Technology, 76131 Karlsruhe, Germany}

\author{{Markus} {Garst}}
\affiliation{Institut für Theoretische Festkörperphysik, Karlsruhe Institute of Technology, 76131 Karlsruhe,  Germany}
\affiliation{Institute for Quantum Materials and Technology, Karlsruhe Institute of Technology, 76131 Karlsruhe, Germany}

\author{{Peter} {Schmitteckert}}
\affiliation{{HQS Quantum Simulations GmbH}, {Rintheimer Stra{\ss}e 23}, {76131 Karlsruhe}, {Germany}}

\date{\today}


\begin{abstract}
    We present important use cases and limitations when considering results
    obtained  from Cluster Perturbation Theory (CPT).
    CPT combines the solutions of small individual clusters of an
    infinite lattice system with the Bloch theory of conventional
    band theory in order to provide an approximation for the Green's
    function in the thermodynamic limit.
    To this end we are investigating single-band and multi-band
    Hubbard models in one- and two-dimensional systems. A special
    interest is taken  in the supposed pseudo gap regime of the
    two-dimensional square lattice at half filling and intermediate interaction
    strength ($U \leq 3t$) as well as the metal-insulator transition.
    We point out that the finite-size level spacing of the cluster
    limits the resolution of spectral features within CPT.
    This restricts the investigation of asymptotic properties of the metal-insulator
    transition, as it would require much larger cluster sizes that are beyond computational
    capabilities.
\end{abstract}


\maketitle
\graphicspath{{./}}

\section{Introduction}\label{sec1}

The Hubbard model probably belongs to the most studied systems in solid state theory.
Although its Hamiltonian possesses a simple form, it captures important
aspects of various many-body phenomena like Mott-insulating states,
antiferromagnetism and superconductivity.%
\cite{reich_heavy-fermion_1988,tremblay_pseudogap_2006,
huscroft_pseudogaps_2001, gull_superconductivity_2013,
Arovas_hubbard_model}.
The Hamiltonian has three terms: the first term describes the hopping of the
electrons on the lattice, the second term a repulsive Coulomb interaction
of spin up and spin down electrons on the same site and the third term is the
chemical potential, which we shifted such that half filling corresponds to $\mu=0$
for bi-partite lattices:
\begin{align}
    {\cal H} &= - \sum_\sigma \sum_{x,y}
                t_{x,y} \hat{c}^\dagger_{x, \sigma} \hat{c}_{y, \sigma}
            + U \sum_x \hat{n}_{x, \uparrow} \hat{n}_{x, \downarrow} \nonumber \\
            &- (\mu + U/2) \sum_x (\hat{n}_{x, \uparrow} + \hat{n}_{x, \downarrow}) ,
\end{align}
where $x$ and $y$ are labelling the lattice sites and $\sigma=\uparrow,\downarrow$
denotes the spin index. With $\hat{c}^\dagger_{x, \sigma},
\hat{c}_{y, \sigma}$ we denote the fermionic creation and annihilation
operators and $\hat{n}_{x,\sigma} = \hat{c}^\dagger_{x, \sigma} \hat{c}^{}_{x, \sigma}$ is
the occupation number operator.

One interesting aspect of the Hubbard model is its Mott-insulating state
at high interaction strength as well as the associated metal-insulator transition
it supposedly captures. In this regard a pseudogap regime at intermediate
interaction strength has been discussed \cite{tremblay_pseudogap_2006, schafer_fate_2015}.
Within this study we investigated
this regime using Cluster Perturbation Theory (CPT).
Introduced by Senechal et al.\ \cite{senechal_spectral_2000} CPT has shown remarkable
results when applied to the Hubbard model,
despite being of low numerical cost. While these results caught our initial interest
for the method, we came to the conclusion that care has to be taken when
interpreting the results of CPT, especially concerning features like spectral gaps.
In the following, we will first outline the method, apply it to the systems of interest
and then analyse carefully the accuracy of the results by comparing the one dimensional
case to exact results using Bethe ansatz.

\section{Methods}\label{sec2}
\subsection{Cluster Green's functions}
In Cluster Perturbation Theory (CPT) the main objective is to construct
an approximation to the retarded Green's function
${\cal G}^r(\textbf{k}, \omega)$ of a given lattice system in the
thermodynamic limit.
This function is especially useful as it provides direct access to the spectral function
\cite{bruus_many-body_2004, fetter2012quantum}:
\begin{align}
    {\cal A}(\textbf{k}, \omega) = -\frac{1}{\pi} \text{Im}
    {\cal G}^r(\textbf{k}, \omega).
\end{align}
As CPT aims at approximating the Green's function of the full system
by combining the solutions of small finite clusters cut out of the infinite
lattice, we first have to discuss how to obtain the 
interacting Green's function on such a cluster. For this we first define
the retarded Green's function for two fermionic operators $\hat{A}$ and $\hat{B}$ as:
\begin{align}
    {\cal{G}}^r_{\hat{A},\hat{B}}(t, t') =
    -i \Theta(t - t') \langle \{\hat{A}(t), \hat{B}(t') \} \rangle
\end{align}
where $\{\ldots, \ldots\}$ is the anticommutator,
$t$ and $t'$ are time arguments and $\Theta(t)$ is the Heavyside Theta function.
Note that we are only interested in the $T=0$ case, which means that
the expectation value ($\langle ... \rangle$) only consists of the ground
state $\ket{\Psi_0}$.
For the retarded Green's function we have $t>t'$ and as the
Hamiltonian of interest is time-independent, we can assume $t'$ to be zero.
As we are going to use a Chebyshev expansion, it is convenient to rewrite
the retarded Green's function in terms of two new functions
${\cal{G}}^+_{\hat{A},\hat{B}}(t)$ and ${\cal{G}}^-_{\hat{B},\hat{A}}(t)$
\cite{schmitteckert_calculating_2010}:
\begin{align}
    {\cal{G}}^r_{\hat{A},\hat{B}}(t) &=
    -i \Theta(t) \langle \{\hat{A}(t), \hat{B}(0) \} \rangle, \nonumber \\
        \nonumber &=  -i \Theta(t) \langle \hat{A}(t) \hat{B}(0) \rangle
            -i \Theta(t) \langle \hat{B}(0) \hat{A}(t) \rangle, \\
        &= {\cal{G}}^+_{\hat{A},\hat{B}}(t) - {\cal{G}}^-_{\hat{B},\hat{A}}(t) .
\end{align}
Where we used the definitions:
\begin{align}
    {\cal{G}}^+_{\hat{A},\hat{B}}(t) &=
    -i \Theta(t) \langle \hat{A}(t) \hat{B}(0) \rangle ,
    \\
    {\cal{G}}^-_{\hat{B},\hat{A}}(t) &=
    i \Theta(t) \langle \hat{B}(0) \hat{A}(t) \rangle  .
\end{align}
Performing a Fourier transformation we can obtain the Green's function
in the frequency domain as:
\begin{align}
    \label{eq:final Gp}
    {\cal{G}}^+_{\hat{A},\hat{B}}(\omega) =
    -\bra{\Psi_0} \hat{A} [{\cal H}-E_0-(\omega + i\eta)]^{-1} \hat{B} \ket{\Psi_0} .
\end{align}
\begin{align}
    \label{eq:final Gm}
    {\cal{G}}^-_{\hat{B},\hat{A}}(\omega) =
    -\bra{\Psi_0} \hat{B} [{\cal H}-E_0+(\omega + i\eta)]^{-1} \hat{A} \ket{\Psi_0} ,
\end{align}
where $\eta >0$ is an infinitesimal parameter that ensures convergence.

\subsection{Chebyshev Expansion}
Expressions like (\ref{eq:final Gp}) and (\ref{eq:final Gm}) can be very efficiently
handled using Chebyshev polynomials \cite{braun_numerical_2014}.
They contain the function
\begin{align}
    f^{\pm}_z (x) = -i \int^{\pm \infty}_0 {\mathrm e}^{i(\pm z - x)t}  \,d t= \frac{1}{\pm z - x} ,
\end{align}
with $x, \text{Re}(z) \in \mathbb{R}$ and $\text{Im}(z) > 0$, that can be expanded
using Chebyshev polynomials of the first kind $T_n(x)$:
\begin{align}
    f^{\pm}_z (x) = \sum_{n=0}^\infty \alpha^{\pm}_{n}(z) T_n(x) ,
\end{align}
with the expansion coefficients:
\begin{align}
    \alpha^{\pm}_{n}(z) =
    \frac{2/(1 + \delta_{n,0})}{(\pm z)^{n+1}(1 + \sqrt{z^2}
    \sqrt{z^2-1}/z^2)^n \sqrt{1-1/z^2}} .
\end{align}
For the polynomials the following recursion relation holds:
\begin{align}
    \ket{\Phi_0} = b \ket{\Psi_0},
\end{align}
\begin{align}
    \ket{\Phi_1} = [a({\cal H}-E_0)-b] \ket{\Phi_0},
\end{align}
\begin{align}
    \ket{\Phi_{n+1}} = 2 [a({\cal H}-E_0)-b] \ket{\Phi_n}-\ket{\Phi_{n-1}},
\end{align}
where we choose the two parameters $a,b \in \mathbb{R}$ to fit the spectrum
of the operator $a({\cal H}-E_0)-b$ into the interval $(-1,1)$, required
by the orthogonality relation of the Chebyshev polynomials. With this
we can identify the Green's functions as:
\begin{align} \label{ChExp}
    {\cal{G}}^{\pm}_{\hat{B},\hat{A}}(\omega) &=
    a \sum_{n=0}^\infty \alpha_n^\pm (\pm a (\omega+ i \eta) - b) \mu_n ,
\end{align}
where the $\mu_n$ are often referred to as Chebyshev moments and are defined
as the expectation values of the polynomials:
\begin{align}
    \mu_n = \bra{\Psi_0} \hat{A} T_n(a({\cal H}-E_0)-b)\hat{B}\ket{\Psi_0}
    = \bra{\Psi_0} \hat{A} \ket{\Phi_n} .
\end{align}
In order to calculate these moments
for the Green's function of the finite cluster,
we require the Hamiltonian in a many
particle basis. To construct the ground state we employ a sparse matrix diagonalization
as for example introduced in Ref. \cite{senechal_introduction_2010}.
The main idea is to explicitly encode how a specific Hamiltonian acts on
basis states in the occupation number representation. For this, one needs to at least encode all the basis states in a particular number
sector. Although this can be efficiently done by saving each basis state as the
bitwise representation of an integer, the computational space still grows exponentially,
making it only usable for very small clusters. Due to limited computational resources
our calculations did not exceed 18 site calculations. Having constructed the Hamiltonian
the groundstate can be calculated using a Lanczos algorithm.

\subsection{Cluster Perturbation Theory}
\label{sec:Cluster Perturbation Theory method}
The goal of Cluster Perturbation Theory (CPT) is to approximate the
Green's function of a particular lattice model in the thermodynamic
limit by combining the Green's functions of small individual clusters,
for example calculated as described in the previous section.
Introductions to this method are presented in Ref.~\cite{senechal_introduction_2010, pavarini_ldadmft_2011}.
The first step is to split the Hamiltonian into two parts:
\begin{align}
	{\cal H} = \sum_\alpha {\cal H}_\alpha^\text{cluster} + {\cal H}^\text{inter}.
\end{align}
In the first part:
\begin{align}
	{\cal H}^\text{cluster}_\alpha &= ({\cal H}^c_0 + {\cal H}^c_U)_\alpha \nonumber
	= -t \sum_\sigma \sum_{x,y \in \gamma^c_\alpha} \hat{c}^{\dagger}_{x, \sigma}
	\hat{c}^{}_{y, \sigma}  \nonumber \\
	&+ U \sum_{x \in \gamma^c_\alpha} \left( \hat{n}_{x, \uparrow} - \frac{1}{2} \right)
	\left( \hat{n}_{x, \downarrow} - \frac{1}{2} \right) ,
\end{align}
one has the full Hubbard model on small, individual
clusters $\gamma^c_\alpha$, each labeled by the index $\alpha$ and
in the second part:
\begin{align}
	{\cal H}^\text{inter} = -t  \sum_\sigma
		\sum_{x \in \gamma^c_\alpha, y \in \gamma^c_\beta}
		\hat{c}^{\dagger}_{x, \sigma}
		\hat{c}_{y, \sigma} ,
\end{align}
we only have the hopping elements between these individual
clusters. Note that due to this splitting, it can be very
useful to describe any lattice site $\textbf{R}_i$ by a
combination of two new vectors:
\begin{align}
	\textbf{R}_i = \textbf{r}_\alpha + \textbf{r}_m ,
\end{align}
where $\textbf{r}_\alpha$ is the position of the individual clusters in
a new superlattice $\Gamma$ and $\textbf{r}_m$ describes the position
of an individual site within a cluster.\\
One can calculate the Green's function for one of these clusters
and use this result for the Green's function of all other clusters due
to the lattice symmetry. We will refer to this Green's function as the cluster
Green's function ${\cal G}^c(\textbf{r}_m, \textbf{r}_n,\omega)$. The main
idea within CPT consists in calculating the self-energy from the cluster Green's
function and use it to construct an approximation for the self-energy
of the full system. We can obtain the cluster self-energy
$\Sigma^c(\textbf{r}_m, \textbf{r}_n, \omega)$ from a Dyson equation:
\begin{align}
	\Sigma^c(\textbf{r}_m, \textbf{r}_n, \omega) &=
	({\cal G}_{0}^c(\textbf{r}_m, \textbf{r}_n,\omega))^{-1} \nonumber \\
	&- ({\cal G}^c(\textbf{r}_m, \textbf{r}_n,\omega))^{-1} ,
\end{align}
where ${\cal G}_{0}^c(\textbf{r}_m, \textbf{r}_n,\omega)$ is the
non-interacting Green's function on the cluster defined as:
\begin{align}
	({\cal G}_{0}^c(\textbf{r}_m, \textbf{r}_n,\omega))^{-1}
	= \omega + i \eta - {\cal H}^c_{0}(r_m, r_n).
\end{align}
Therefore we obtain for a particular entry of the system self-energy
$\Sigma^s(\textbf{R}_i, \textbf{R}_j,\omega)$ in real space, connecting
two sites on the same cluster:
\begin{align}
	\Sigma^s(\textbf{R}_i, \textbf{R}_j,\omega)
	&= \Sigma^s(\textbf{r}_\alpha + \textbf{r}_m, \textbf{r}_\alpha
	+ \textbf{r}_n,\omega) \nonumber \\
	&= \Sigma^c(\textbf{r}_m, \textbf{r}_n,\omega),
\end{align}
and all entries of the self-energy connecting sites on different
clusters are set to zero.\\
Finally, we can use this approximation of the self energy,
namely using the cluster self energy for the self energy of the full system,
in a Dyson equation as before, to obtain the
Green's function of the full system:
\begin{align}
	({\cal G}_0^s(\omega))^{-1}
	= \omega + i \eta - {\cal H}^\text{inter} - \sum_{\alpha} {\cal H}^c_{0, \alpha} .
\end{align}
Note that in this way, we treat the non-interacting part exactly.
This is why one should view the CPT approximation as a perturbation theory in $U$
rather than a perturbation in the inter-cluster hopping.
Finally we end up with the following expression for the Green's function
of the full system:
\begin{align}
	{\cal G}^s(\textbf{R}_i, \textbf{R}_j, \omega)
	&= (({\cal G}_{0}^s(\textbf{R}_i, \textbf{R}_j, \omega))^{-1} \nonumber \\
		&- \Sigma^s(\textbf{R}_i, \textbf{R}_j, \omega))^{-1}
\end{align}
\subsection{Periodization}
\label{sec:Periodization}

While the just described procedure works for finite systems, it is
important to note that there are also so called periodization schemes,
which allow to extend these results to infinite systems. Here we are going
to use the so called G-scheme, as discussed in Ref. \cite{avella_strongly_2012}.
The main idea is based on arranging clusters in an
infinite superlattice and exploiting its translational symmetry. As
pointed out before, one can split any lattice vector into one
vector defined on the superlattice and one on a cluster.
Therefore, one can similarly split any wave vector $\textbf{k}$ of the 1st BZ into
a combination of a wave vector in a reduced BZ $\tilde{\textbf{k}}$ associated with
the superlattice and one of the Brillouin zone of a single cluster $\textbf{K}$.
This also allows one to split the Fourier transformations into two parts,
one for the cluster and one for the superlattice. Using Bloch's theorem for the
superlattice, one ends up with the following form for a periodized Green's function:
\begin{align}
	\label{eq:periodized GF}
	{\cal G}(\textbf{k}, \omega) = \frac{1}{L} \sum_{a,b}
		e^{-i \textbf{k}(\textbf{r}_a - \textbf{r}_b)}
		{\cal G}_{a,b}(\tilde{\textbf{k}}, \omega) .
\end{align}

\section{Results}
\label{Results}
\begin{figure}
    \centering
    \includegraphics[width=0.45\textwidth]{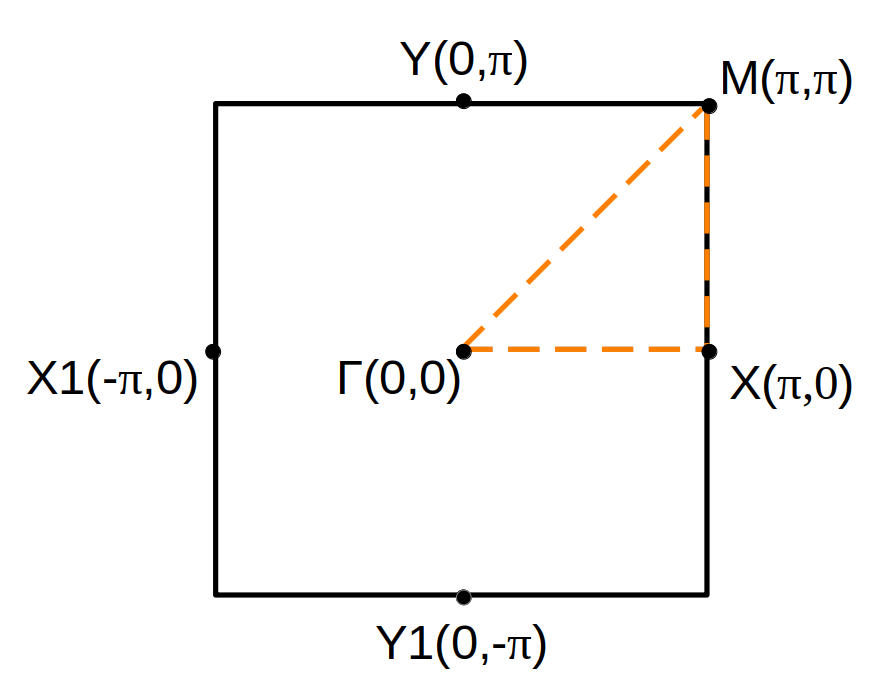}
    \caption{First Brillouin zone (1. BZ) of the 2D square lattice with symmetry points and
    the k-path for the bandstructure plots.}
    \label{symmetrypoints}
\end{figure}
\begin{figure}
    \includegraphics[width=0.45\textwidth]{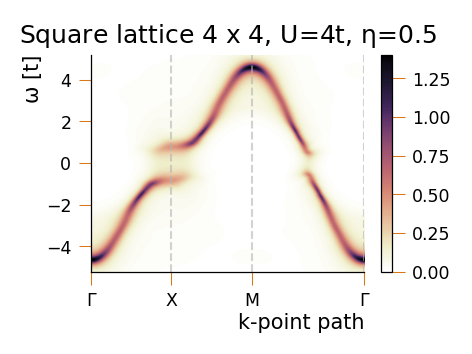}
    \caption{Spectral function of the Hubbard model with U=4t on a 2D square lattice,
    plotted along the high symmetry axis of the 1. BZ. The broadening parameter was
    chosen as $\eta=0.5$. The cluster calculations were performed on a 4x4 cluster with
    120 Chebyshev moments.}
    \label{fig:direct 4x4 U4 eta0.5 pseudogap band structure}
\end{figure}
\begin{figure}
    \includegraphics[width=0.45\textwidth]{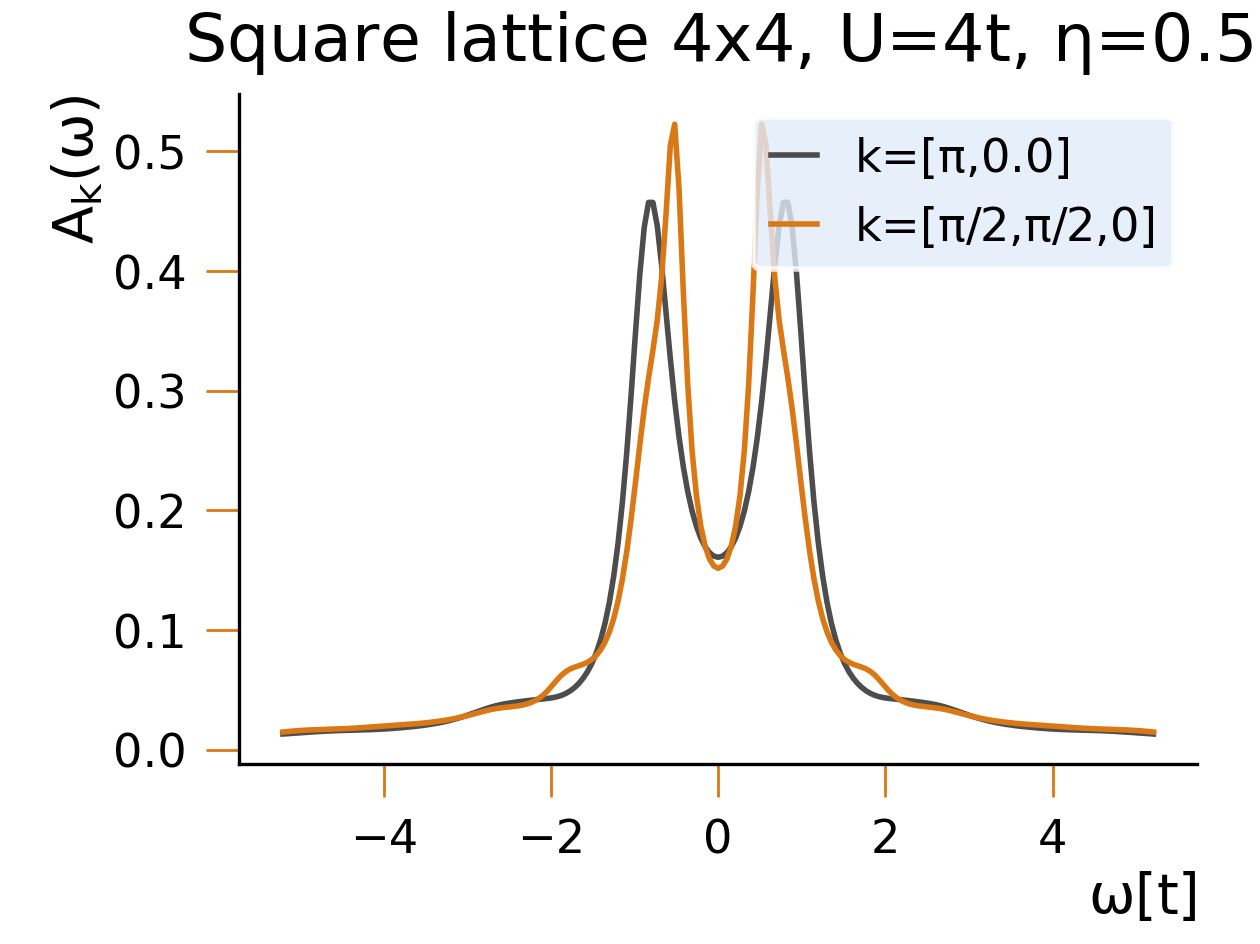}
    \caption{Same spectral function as in Fig. \ref{fig:direct 4x4 U4 eta0.5 pseudogap band structure},
    plotted only at the $X$ point ($k=[\pi, 0, 0]$) and the midpoint between $\Gamma$ and $M$
    point ($k = [\pi/2, \pi/2, 0]$). We can still see a significant spectral weight at $\omega = 0$
    indicating a pseudogap.}
    \label{fig:direct 4x4 U4 eta0.5 pseudogap line plot}
\end{figure}
We can now use the just presented methods to calculate the spectral function
for our main system of interest, the 2D Hubbard model on a square lattice at half
filling (Fig.~\ref{symmetrypoints}, ~\ref{fig:direct 4x4 U4 eta0.5 pseudogap band structure},
~\ref{fig:direct 4x4 U4 eta0.5 pseudogap line plot}). As we can clearly see when considering
the spectral function at the $X$ point and in the midpoint between $\Gamma$ and $M$, CPT
here suggests a considerable spectral weight in the gap at U=4t.
However, we are going to show that there are two parameters that pose large additional
constraints on the resolution that can actually be achieved using CPT. The first parameter
is the convergence aiding factor $\eta$ and the second is the finite cluster size.
While the artefacts induced by the convergence aiding factor are related to our approximation of using only a finite number of Chebyshev moments,
the constraints imposed by the finite cluster size are an inherent limitation
of CPT.
These additional constraints are typically not discussed in detail in the literature,
but as we are going to show they actually prohibit us from making accurate judgements
about the pseudogap at intermediate interaction strengths.
To see this, we will concentrate on the 1D Hubbard model since on the one hand it allows us
to compare to exact results from Bethe ansatz and on the other hand it gives a higher resolution
in k-space.

\subsection{Convergence Aiding Factor}
\label{sec:eta}
\begin{figure}[t] 
    \centering
    \includegraphics[width=0.45\textwidth]{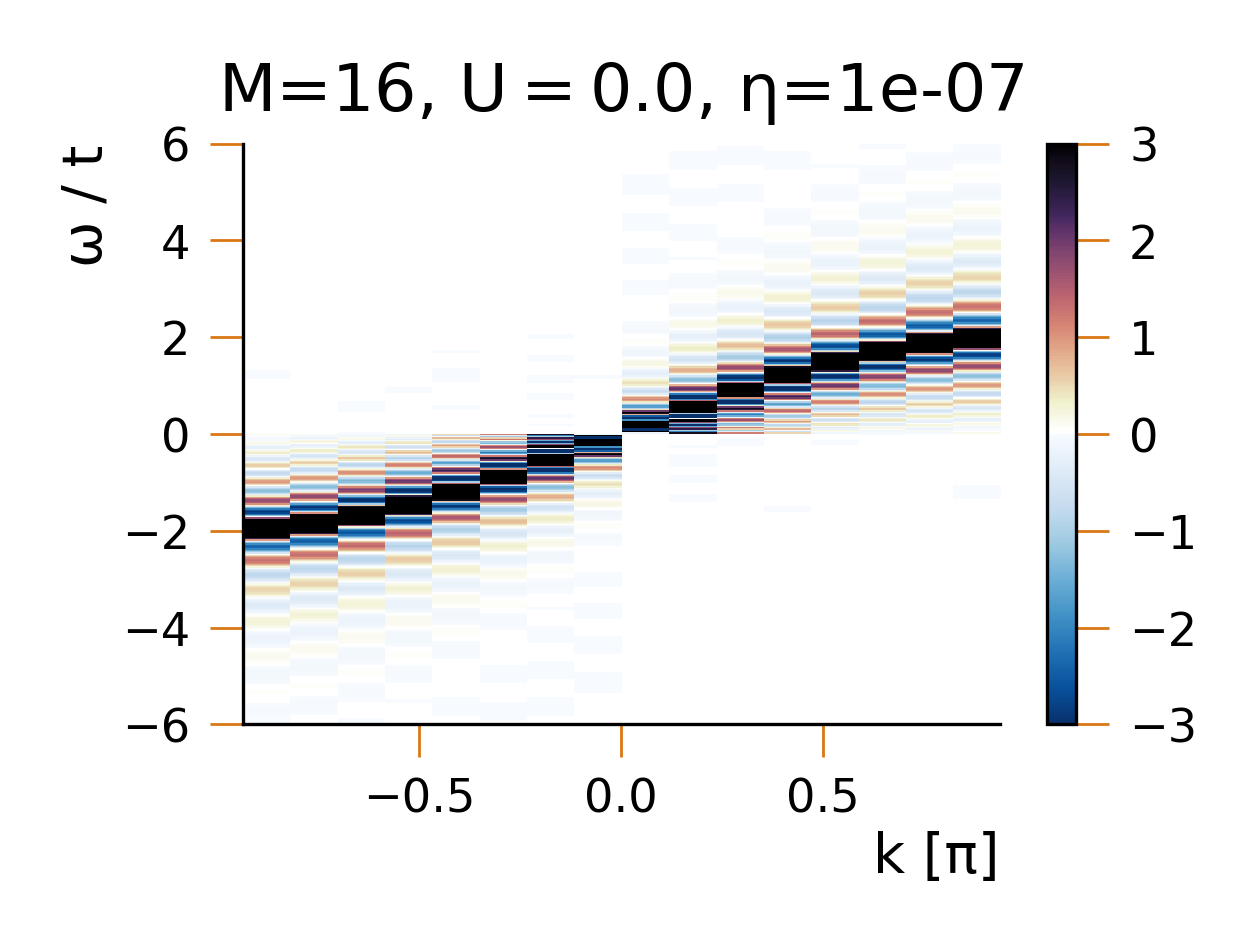}
    \caption{Spectral function obtained from a cluster Green's function of a 16 site tight binding
    chain with a broadening parameter of $10^{-7}$ leading to negative values in the
    spectral function.}
    \label{fig: why eta0 U2}
\end{figure}

\begin{figure} 
    \centering
    \includegraphics[width=0.45\textwidth]{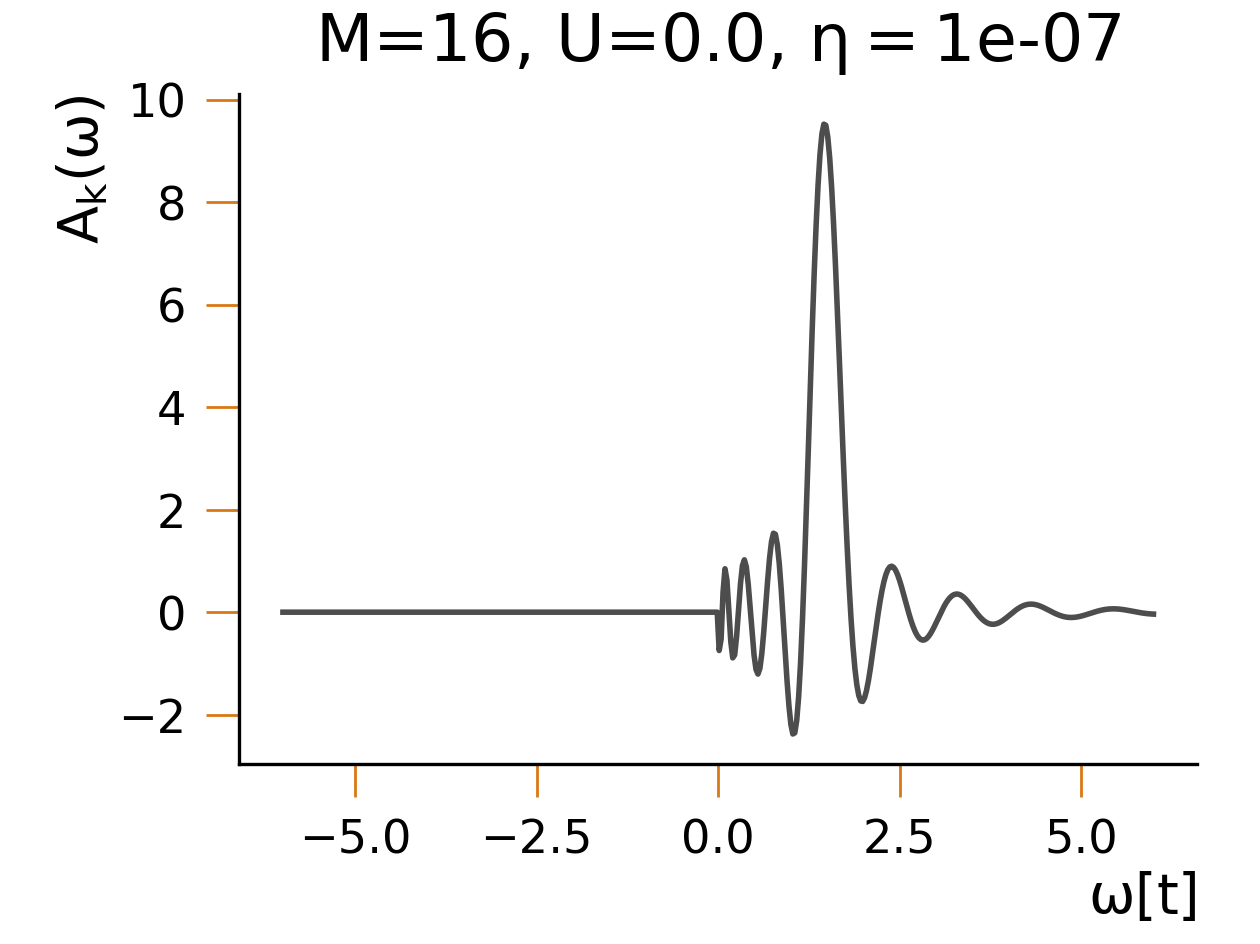}
    \caption{Spectral function of a 16 site tight binding chain at $k=3\pi/4$ with $n_{ch}=60$
    chebyshev moments and a broadening parameter of $10^{-7}$. We can see oscillations around
    the peak leading to negative values in the spectral function.}
    \label{fig: why eta0 U2 lineplot 60}
\end{figure}

\begin{figure} 
    \centering
    \includegraphics[width=0.45\textwidth]{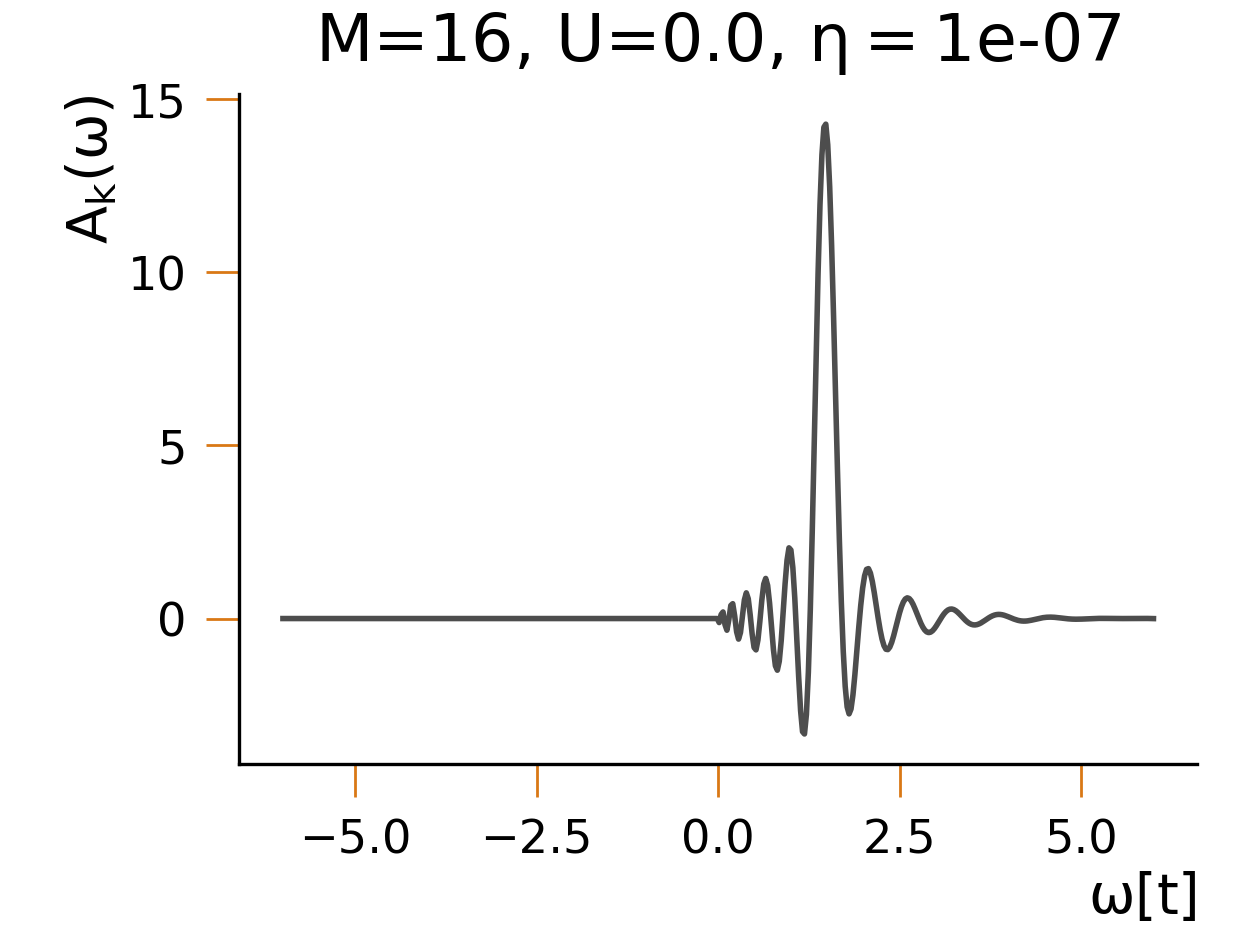}
    \caption{Same spectral function as in Fig. \ref{fig: why eta0 U2} but calculated using
    $n_{ch}=120$ Chebyshev moments. Note that the amplitude of the oscillations does not
    decrease, but their frequency increases in accordance with Gibbs phenomenon.}
    \label{fig: why eta0 U2 lineplot 120}
\end{figure}

The convergence aiding factor $\eta$ enters the Green's function in Eqs.~\eqref{eq:final Gp} and \eqref{eq:final Gm}.
We calculate these Green's functions with the help of the Chebyshev expansion \eqref{ChExp} that, in practise, is
evaluated only with a finite number of Chebyshev moments. This leads to artefacts in the spectral function in the form of Gibbs oscillations.
This is illustrated for the spectral function of the non-interaction 1D chain evaluated for a cluster with 16 sites,
see Fig.~\ref{fig: why eta0 U2}. In Fig.~\ref{fig: why eta0 U2 lineplot 60} we show the spectral function for a specific
$\textbf{k}$-value as a function of frequency that clearly displays oscillations. Note that a higher Chebyshev order only
increases the frequency of these oscillations
(see Fig.~\ref{fig: why eta0 U2 lineplot 120}) but does not change their
magnitude. These oscillations can be identified as Gibbs oscillations that usually arise when approximating a
sharp step function (in this case the $\delta$-peak) by a finite Fourier expansion series. In order to suppress these artificial
oscillations we will choose a sufficiently large value for the broadening parameter $\eta$, such that the
Chebyshev expansion is capable of resolving the peak without Gibbs
oscillations.
In addition, the finite cluster is naturally
characterized by a finite level spacing. In order to
mimic an infinite system and to obtain smooth
bands, a finite broadening parameter $\eta$ has to be chosen
such that it smears out the effect of the
finite level spacing \cite{schmitteckert_calculating_2010}.
\begin{figure} 
    \centering
    \includegraphics[width=0.45\textwidth]{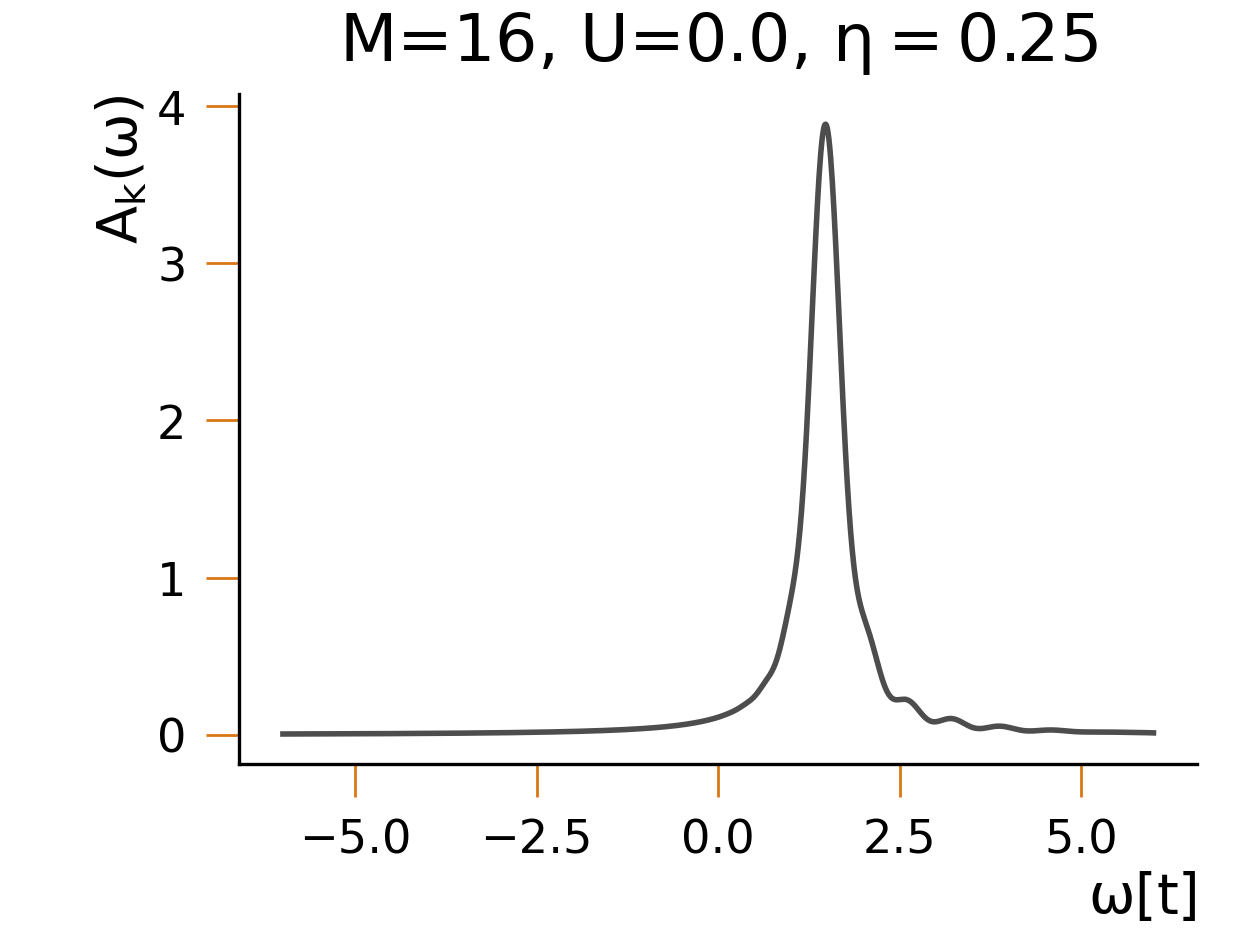}
    \caption{Again the same spectral function as in Fig. \ref{fig: why eta0 U2} calculated using
    $n_{ch}=120$ Chebyshev moments but now with a broadening of $\eta = 0.25$. The oscillations
    disappear, because the peak is artificially broadened.}
    \label{fig: why eta0.25 U2 lineplot}
\end{figure}
\begin{figure} 
    \centering
    \includegraphics[width=0.45\textwidth]{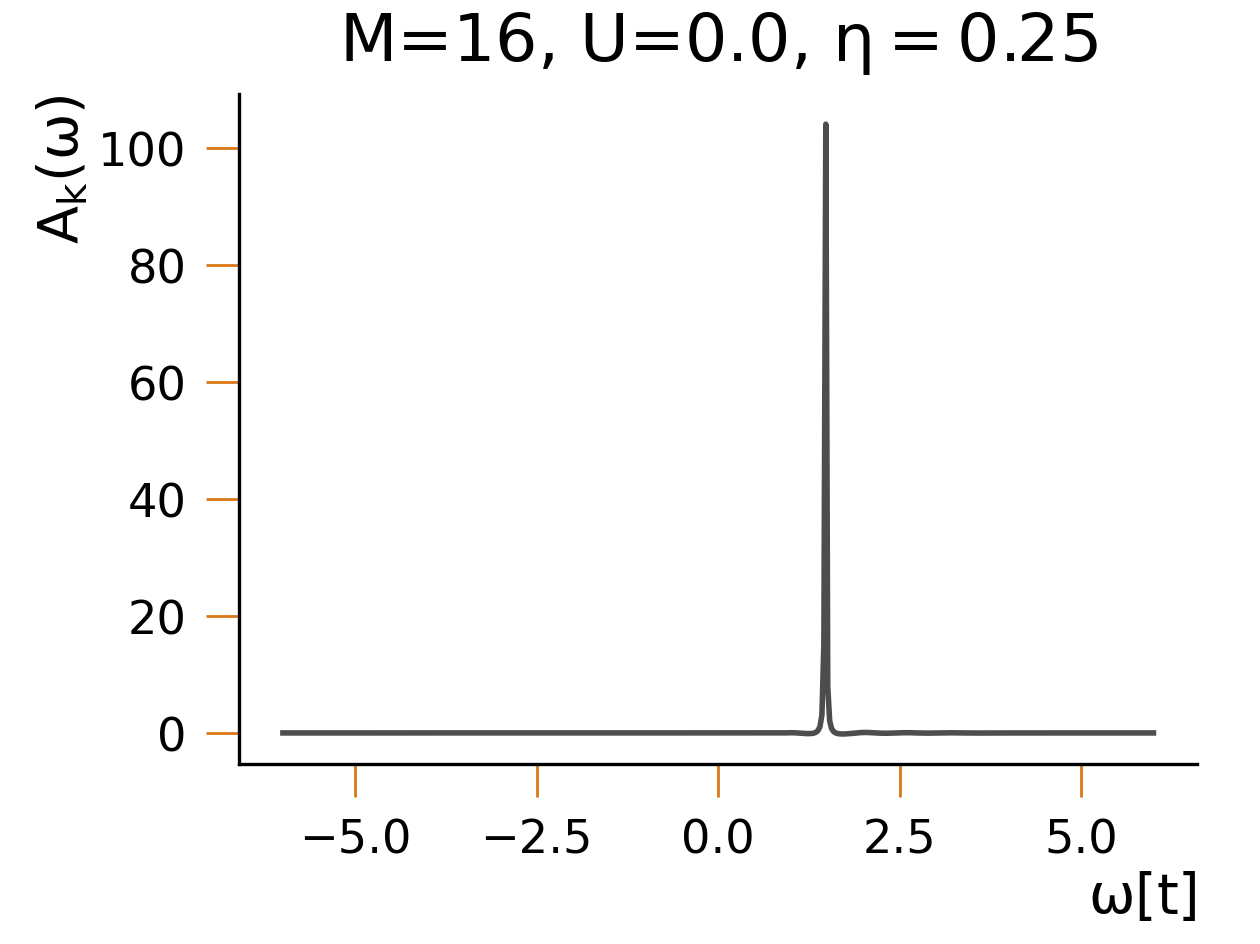}
    \caption{Removing the artificial broadening by means of eq.~\eqref{eq:fix broadening} results in a
    sharp $\delta$-like peak in the spectrum.}
    \label{fig: why eta0.25 U2 lineplot cheby120 eta0.25}
\end{figure}

While this means that altogether one has to choose $\eta$ rather large
(Fig. \ref{fig: why eta0.25 U2 lineplot}), one should also realize that one
can counteract the effects of this broadening to a large extend by including
the same large parameter for $\eta$ in the non interacting Green's function
when calculating the self energy. Effectively this corresponds to subtracting
$\eta$ from the inverse of the Green's function,
\begin{align}
    \label{eq:fix broadening}
	{\cal G}^c(\omega) = \left[ ( {\cal G}^c(\omega, \eta_C) )^{-1} - i\eta_C \right]^{-1}
\end{align}
While this procedure works extremely well, as shown in
Fig.~\ref{fig: why eta0.25 U2 lineplot cheby120 eta0.25}, it still
depends on the exact choice for $\eta$, and it is {\it a priori} unclear which value to choose for $\eta$.
Here we want to propose two different approaches.

The first approach uses the typical single particle level spacing
of the non interacting cluster for the broadening parameter $\eta$,
that can be estimated as
\begin{align}
    \eta = \frac{4t}{M_C}
\end{align}
where $4t$ is the bandwidth with $t$ the hopping amplitude and $M_C$ the cluster size.
This smears the discrete cluster levels (Fig.~\ref{fig:clusterGF U0 small eta}),
and leads to a good approximation of the continuous cosine band structure one expects for a 1D system (Fig.~\ref{fig:clusterGF U0 larger eta}).
Applying this choice for the CPT approximation to the 1D Hubbard model,
one finds a finite spectral weight within the Hubbard gap
illustrating the numerical artefact that is induced by a large $\eta$,
see Figs.~\ref{fig:eta0.5 direct U2} and \ref{fig:eta0.5 direct U2 line plots}.
{ 
\begin{figure}[h!]
    \centering
    \includegraphics[width=0.45\textwidth]{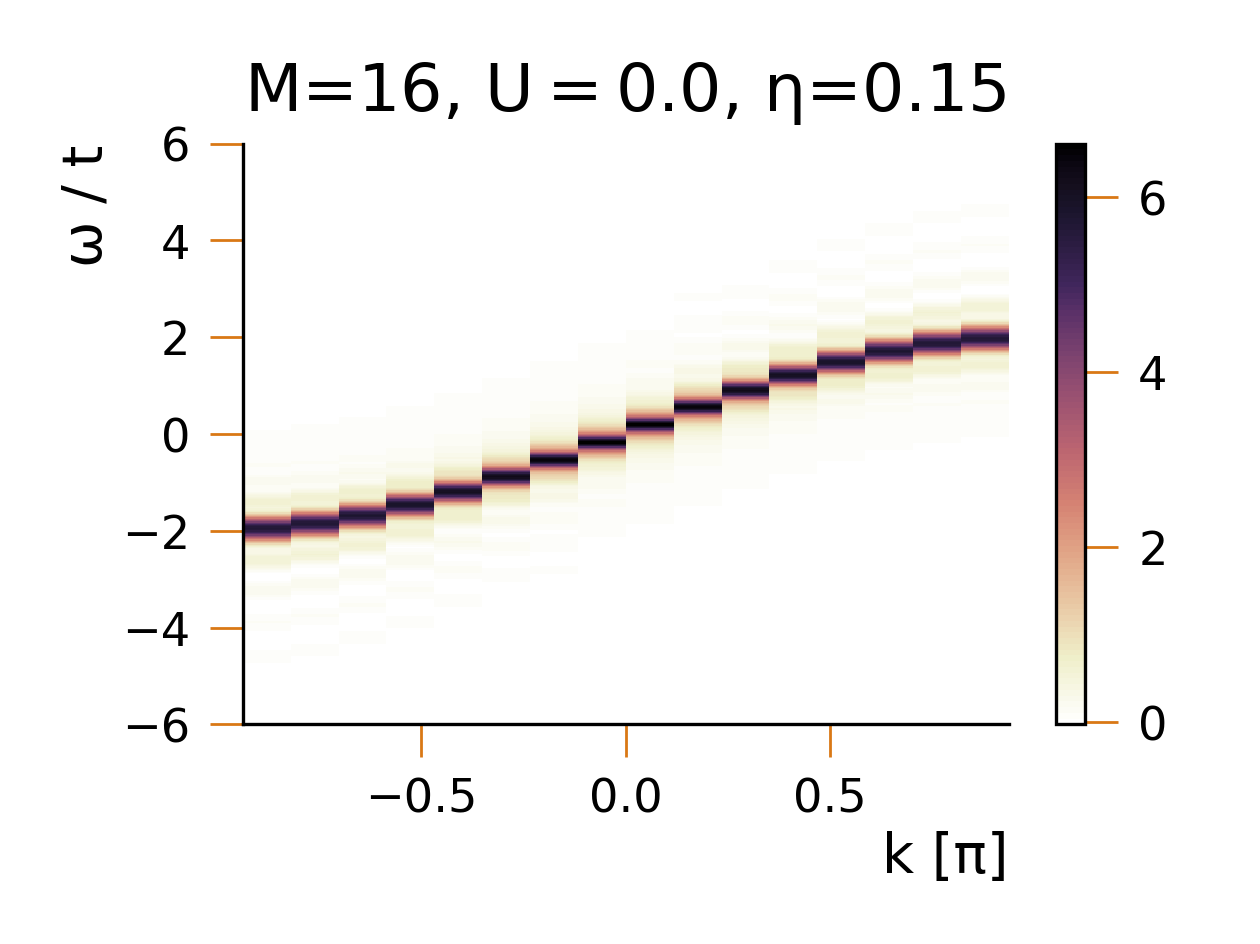}
    \caption{Spectral function of a 16 site Hubbard model at U=0 (tight binding chain) with a broadening
    of $\eta=0.15$. We can see individual levels of the cluster.}
    \label{fig:clusterGF U0 small eta}
\end{figure}
\centering
\begin{figure}[h!]
    \centering
    \includegraphics[width=0.45\textwidth]{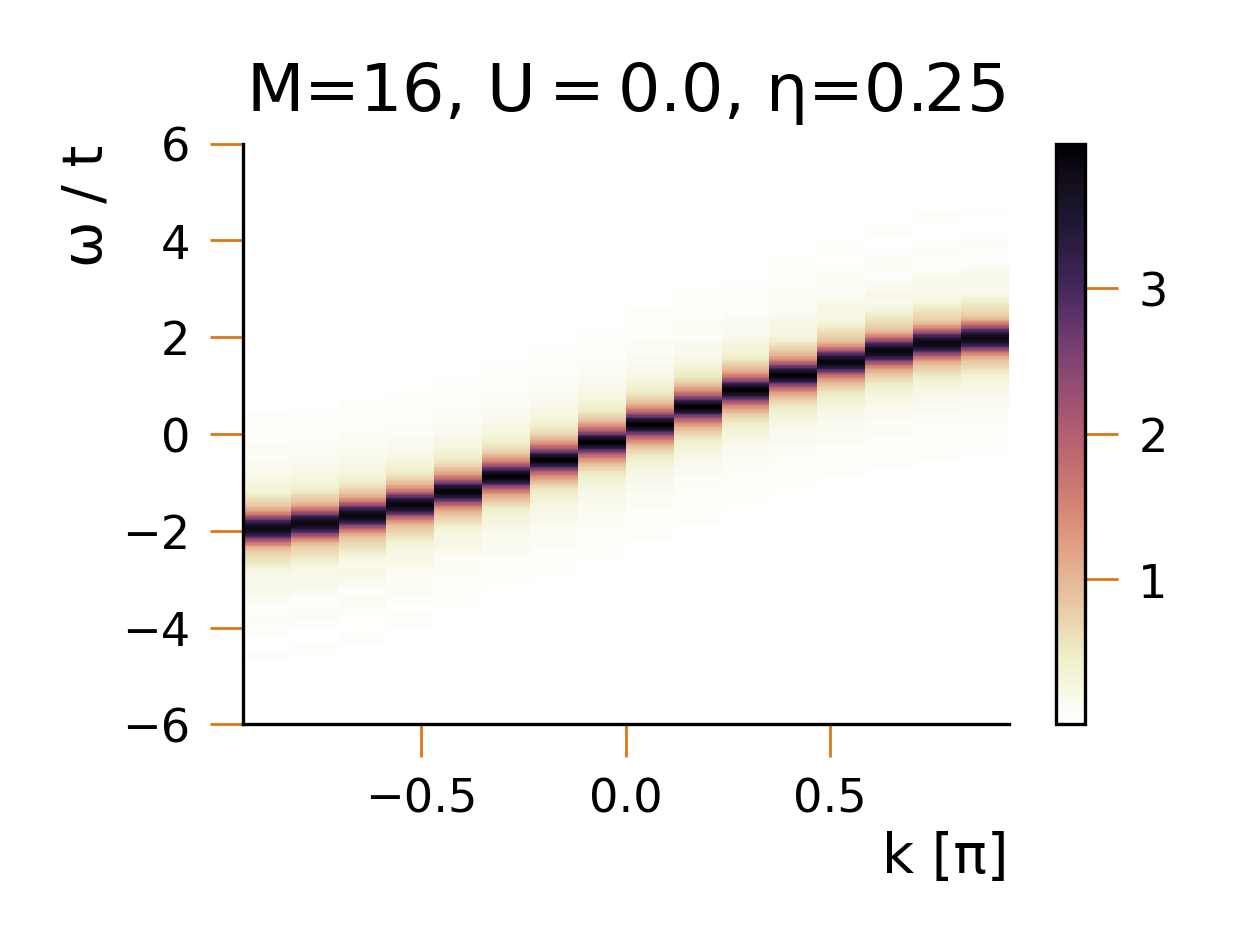}
    \caption{Spectral function for the same system as in Fig. \ref{fig:clusterGF U0 small eta} but with a
    broadening of $\eta=0.25$. We can see that the peaks start to overlap and to resemble a continuous
    cosine bandstructure, as expected for an infinite tight binding chain.}
    \label{fig:clusterGF U0 larger eta}
\end{figure}
\begin{figure}[h!]
    \centering
    \includegraphics[width=0.45\textwidth]{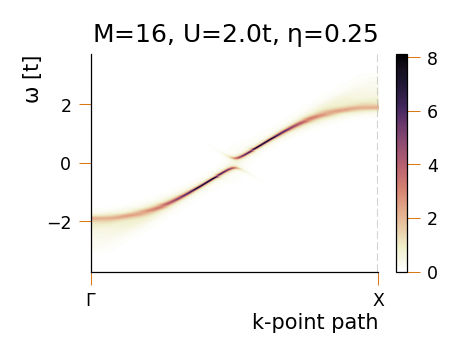}
    \caption{CPT result for a 1D Hubbard model at U = 2t based on a 16 site cluster with broadening
    $\eta=0.25$, based on the average single particle level splitting. We obtain two continuous
    bandstructures with the expected Hubbard gap.}
    \label{fig:eta0.5 direct U2}
\end{figure}
\begin{figure}[h!]
    \centering
    \includegraphics[width=0.45\textwidth]{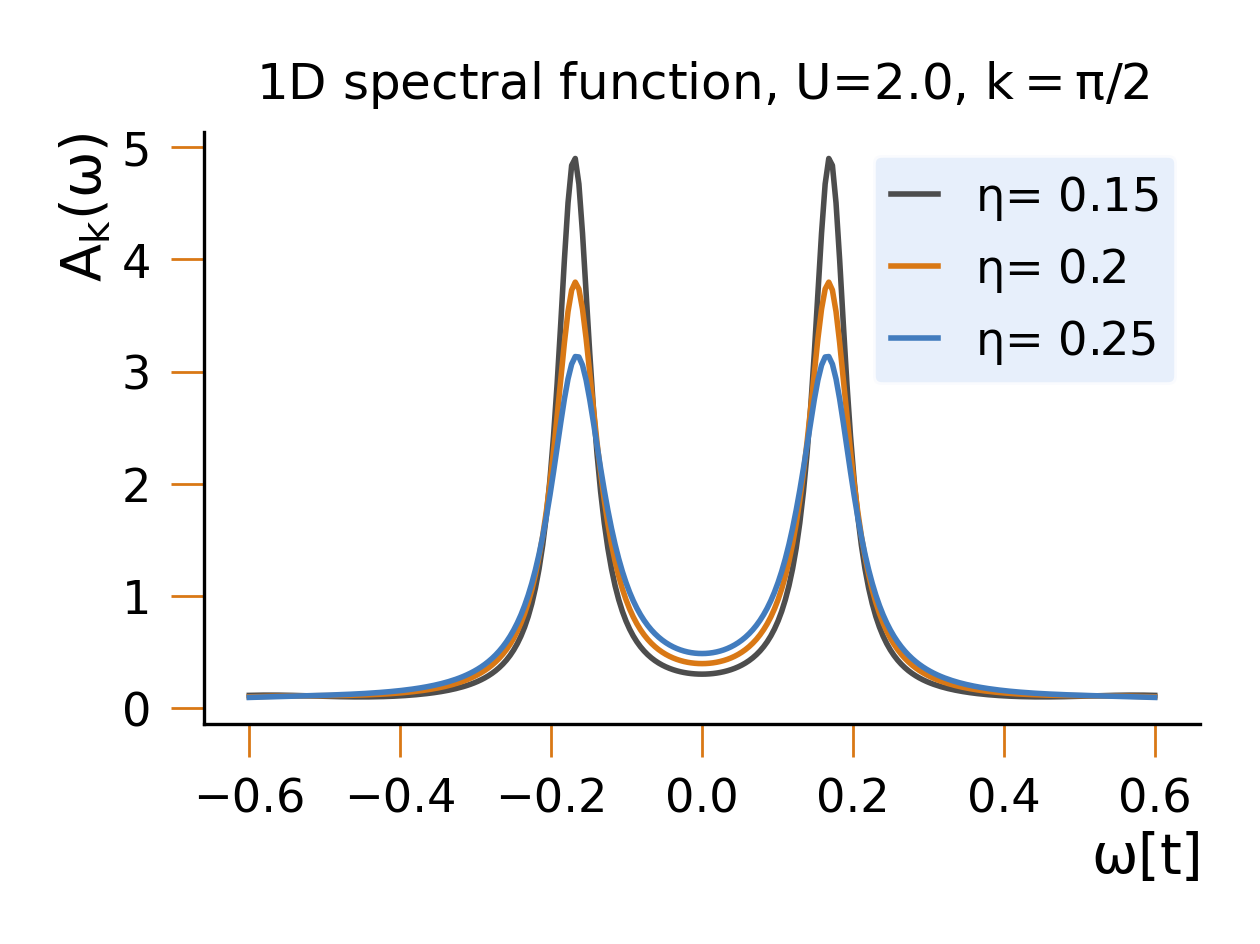}
    \caption{CPT result for the same system as in Fig. \ref{fig:eta0.5 direct U2} shown only
    for $\bold{k}=\pi/2$ but with different broadening parameters. We can see, that the residual spectral
    weight in the gap strongly depends on the broadening parameter.
    }
    \label{fig:eta0.5 direct U2 line plots}
\end{figure}
}

\begin{figure}[h!]
    \centering
    \includegraphics[width=0.45\textwidth]{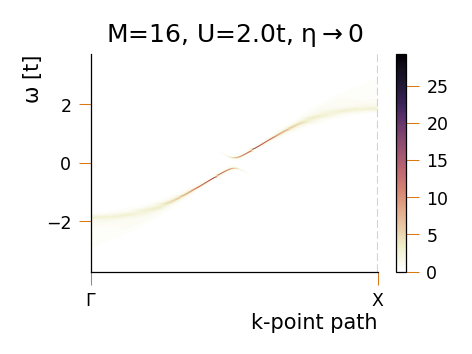}
    \caption{Again CPT results for the same system as in Fig. \ref{fig:eta0.5 direct U2} but
    using an extrapolation scheme for the broadening parameter. We again obtain two continuous bands
    separated by the Hubbard gap, however the bands are narrower.}
    \label{fig:extrapolated eta direct}
\end{figure}

\begin{figure}[h!]
    \centering
    \includegraphics[width=0.45\textwidth]{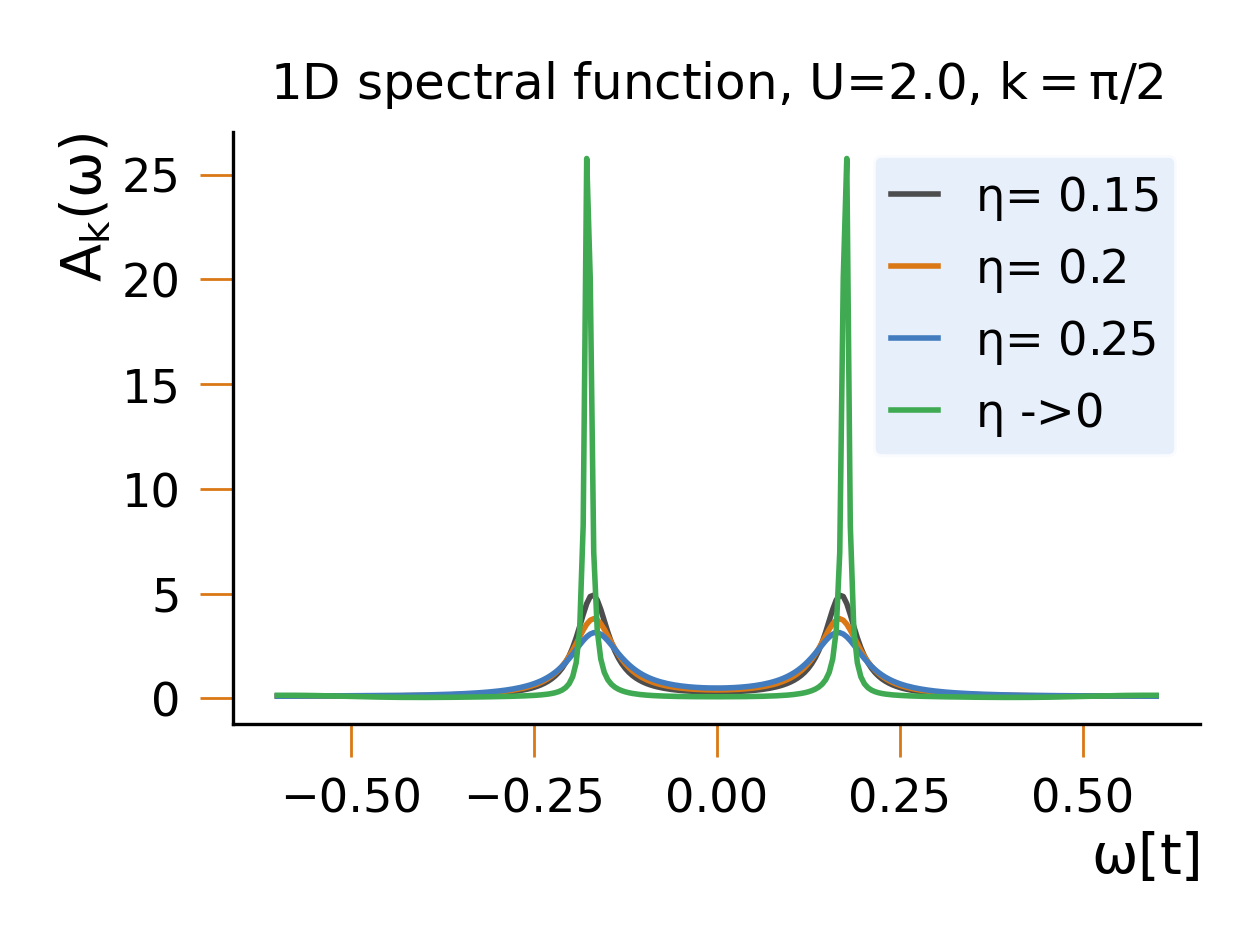}
    \caption{Same plot as in Fig. \ref{fig:eta0.5 direct U2 line plots} but now for the CPT data obtained
    with the extrapolation scheme for the broadening parameter. We can see that the spectral function in
    the gap almost disappears.}
    \label{fig:line plot U2 different eta}
\end{figure}

As a second approach we propose an extrapolation scheme that calculates
the CPT Green's function for multiple values of $\eta$ and performs
an extrapolation of the results to $\eta = 0$
(see Figs.~\ref{fig:extrapolated eta direct} and \ref{fig:line plot U2 different eta}).
Although this procedure is physically sound and results in sharp peaks,
there is no guarantee that this procedure will lead to the correct thermodynamic limit.
In addition, obtaining a sharper peak
does not automatically provide a more accurate result.
Only in cases where
the actual width of the peak is resolved by the Chebyshev expansion,
i.e.\ wider than the many particle bandwidth divided by the number of Chebyshev moments,
the extrapolated peaks could be considered reliable.
Remarkably, despite the peak height and width being $\eta$ dependent,
the peak position appears to be rather stable against a variation of $\eta$.

\subsection{Accuracy of CPT results}
\label{sec:Accuracy}
It is important to note that the finite size level spacing of the cluster used within CPT
acts as a cutoff that limits any spectral resolution. Spectral features like gaps can
only be resolved reliably if they are larger than this cutoff.
As far as interaction effects are concerned, CPT only accurately reflects the same
information that a careful analysis of the cluster result would also provide.

\begin{figure}[h!]
    \centering
    \includegraphics[width=0.45\textwidth]{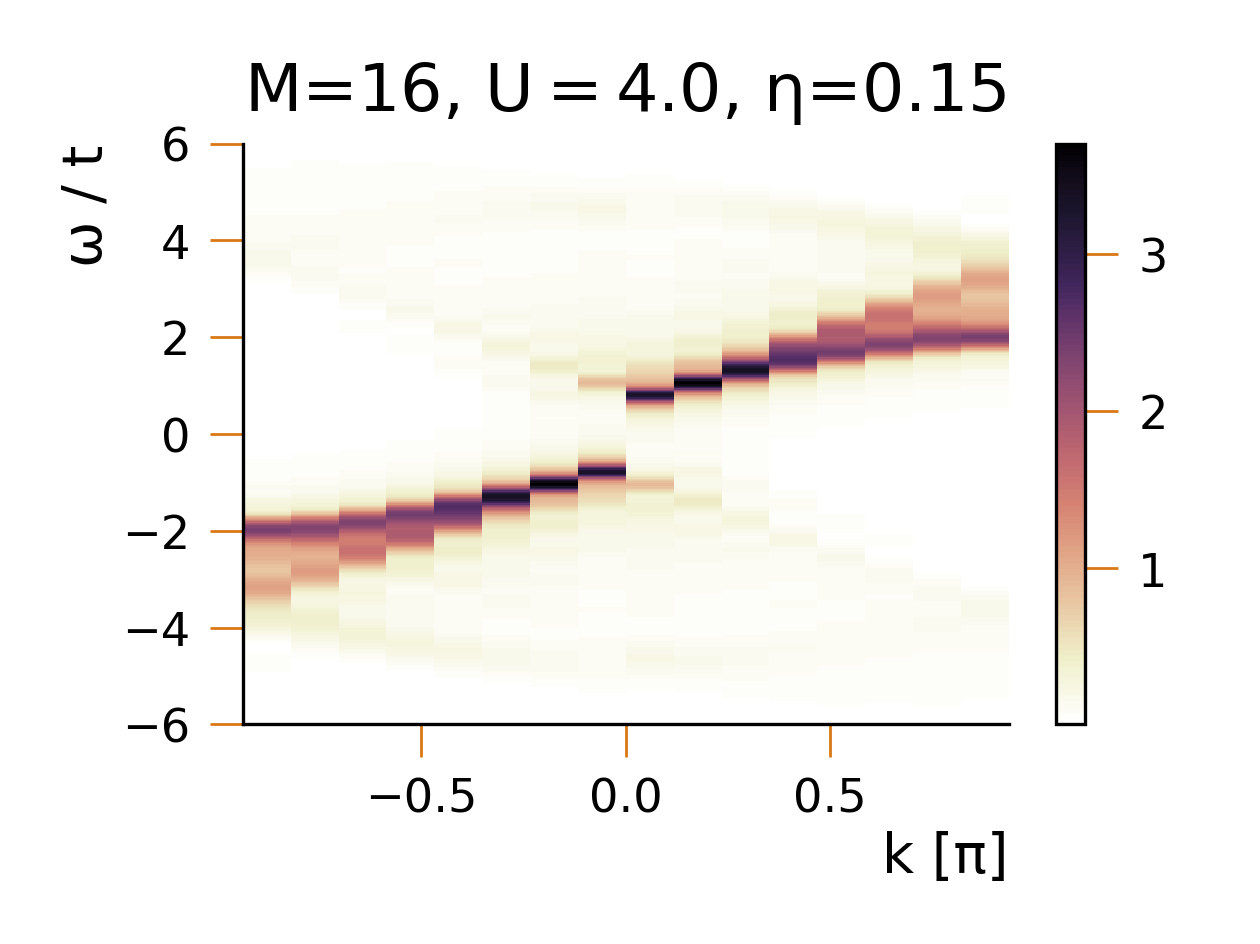}
    \caption{Spectral function of a 16 site Hubbard chain at U=4t and a broadening parameter of
    $\eta=0.15$. Comparing to the non interacting result in Fig. ~\ref{fig:clusterGF U0 small eta} we can see shifts in
    the positions of the single particle levels.}
    \label{fig:clusterGF U4}
\end{figure}

\begin{figure}[h!]
    \centering
    \includegraphics[width=0.45\textwidth]{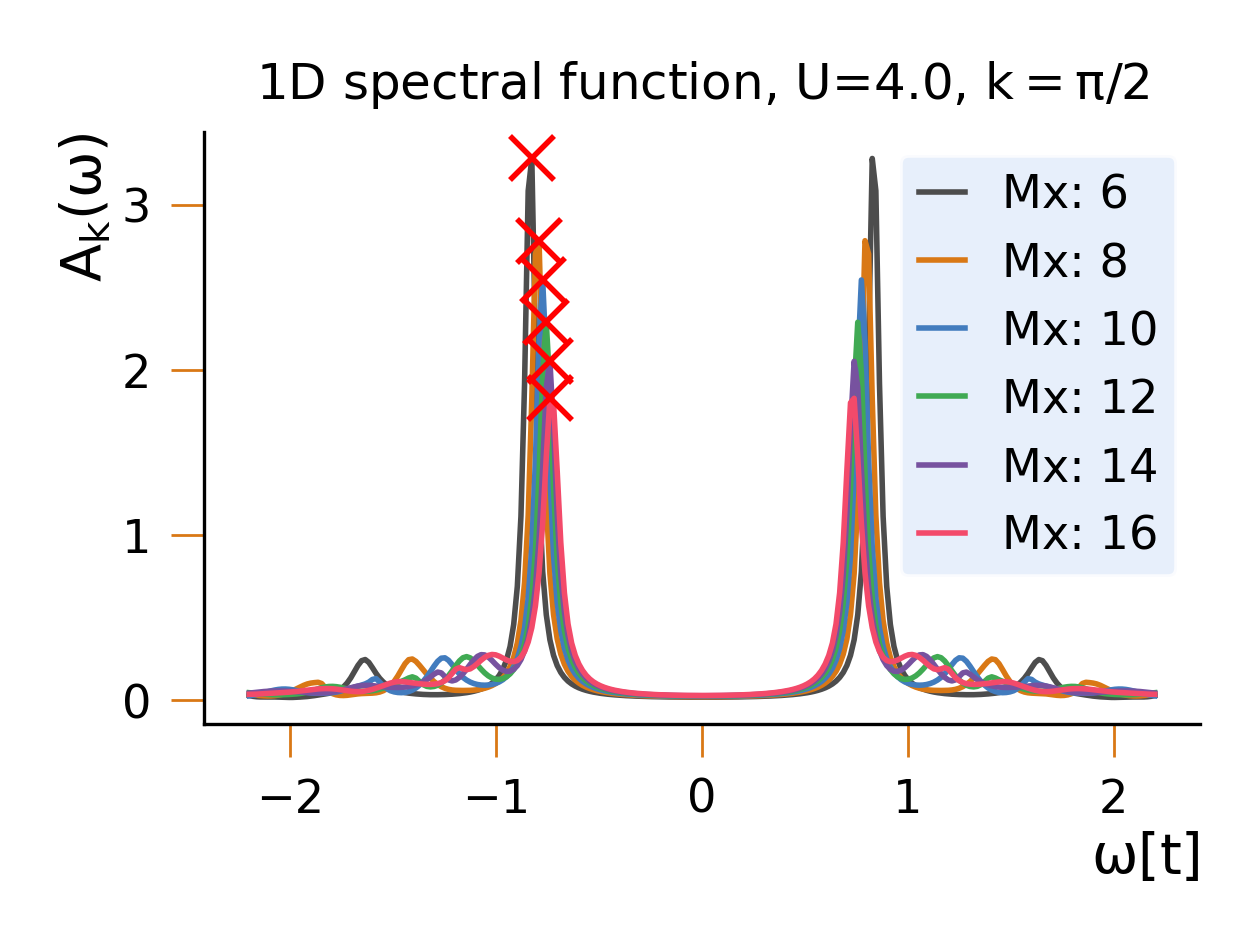}
    \caption{Shown is the influence of the cluster size on the band gap for $U/t=4$. We show
    the spectral function at $k=\pi/2$. The peaks defining the gap get broader and move
    closer together the larger the cluster. Note that the minimum spectral weight within
    the gap stays almost constant with system size and is almost zero due to our
    choice of $\eta=0.15$. The peaks which are marked are used for a finite size
    extrapolation in Fig. \ref{fig:finite size extrapolation U4}.}
    \label{fig:cluster size first order U4}
\end{figure}

\begin{figure}[h!]
    \centering
    \includegraphics[width=0.45\textwidth]{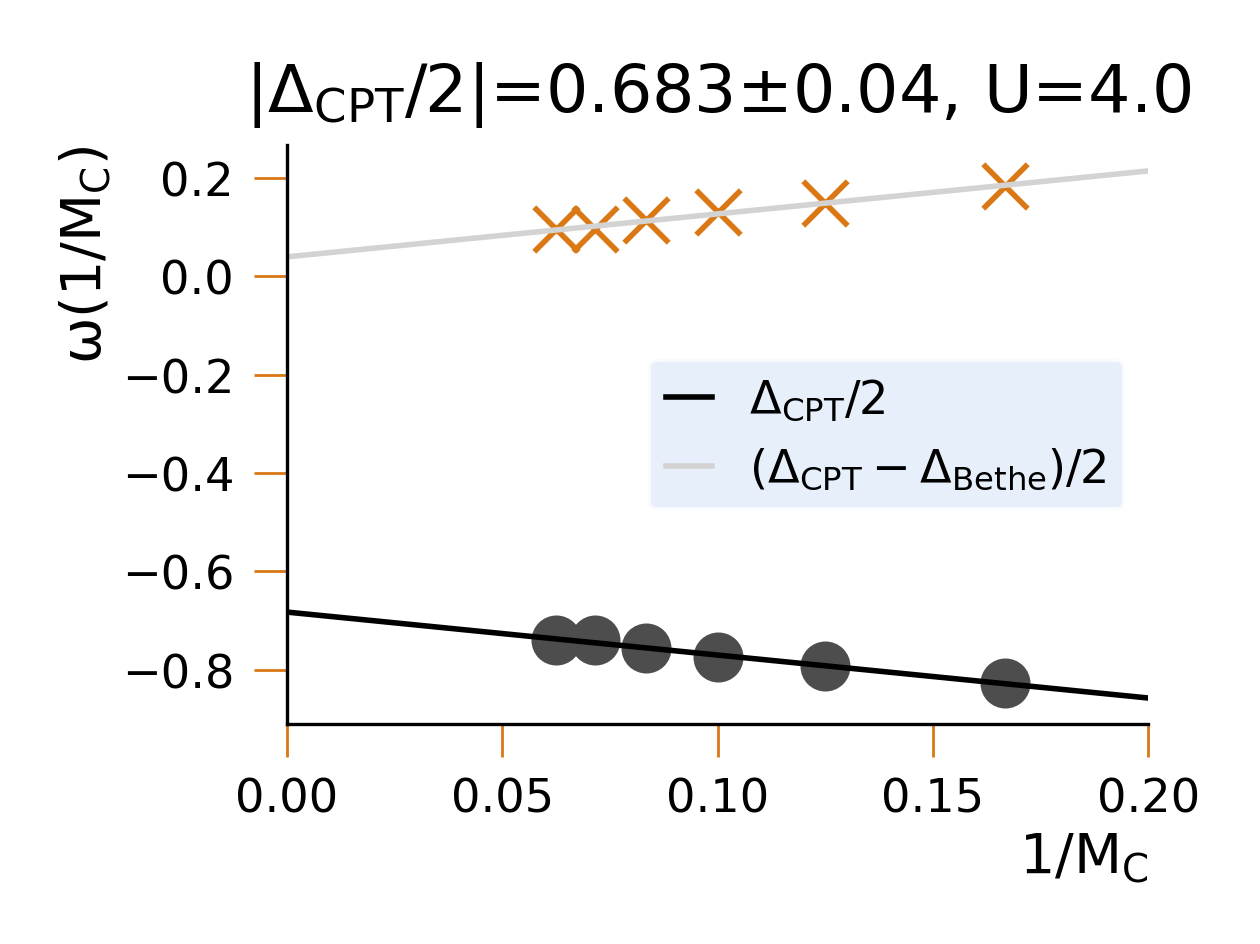}
    \caption{The $\omega$ values of the peaks marked in Fig. \ref{fig:cluster size first order U4}
    are plotted against the inverse of the corresponding cluster size. Doing so allows for a
    finite size extrapolation to zero, corresponding to an infinite cluster size. The
    arrangement of the data suggests a linear extrapolation, which we performed and the result
    at zero is shown in the title. It corresponds to half the bandgap ($\Delta_{CPT}/2$).
    The second value was obtained, by doing the same extrapolation for the deviation from
    the exact Bethe result ($\Delta_{CPT}/2 - \Delta_{Bethe}/2$).}
    \label{fig:finite size extrapolation U4}
\end{figure}

To extract such an information from the cluster result directly, let us consider
the spectral function of a 16 site 1D Hubbard chain without interaction
as was shown in Fig. ~\ref{fig:clusterGF U0 small eta}
and with an interaction of U=4t (Fig.~\ref{fig:clusterGF U4}).
As one would expect for the non interacting case we can see 16 individual levels
forming a cosine shaped band, while for the interacting case the levels in the middle
of the spectrum move apart. Hence, the actual band gap is given by the
shift of the energy levels rather than their frequency difference directly.
Table \ref{tab: Band gap 1D cluster eta corrected} shows the frequency difference of a particular interaction
strength subtracted by the difference in the non interacting case, while the
table below shows the exact results one would get in the thermodynamic limit
using Bethe ansatz (Tab.~\ref{tab: Band gap 1D}).
As we can see, only for $U=4t, 8t$ the gap size agrees up to the second decimal place
with the exact result. However, this means that for $U=1t, 2t$ and using CPT with the currently computationally accessible cluster sizes
we can not judge if there is an actual gap in the system, as the deviation
is on the same order of magnitude as the actual gap. Additionally we can see
that the gap size does not change significantly with the cluster size. Hence
we have to assume that the considered cluster sizes are simply to small to resolve
the gap accurately.
\begin{figure}[h!]
    \centering
    \includegraphics[width=0.45\textwidth]{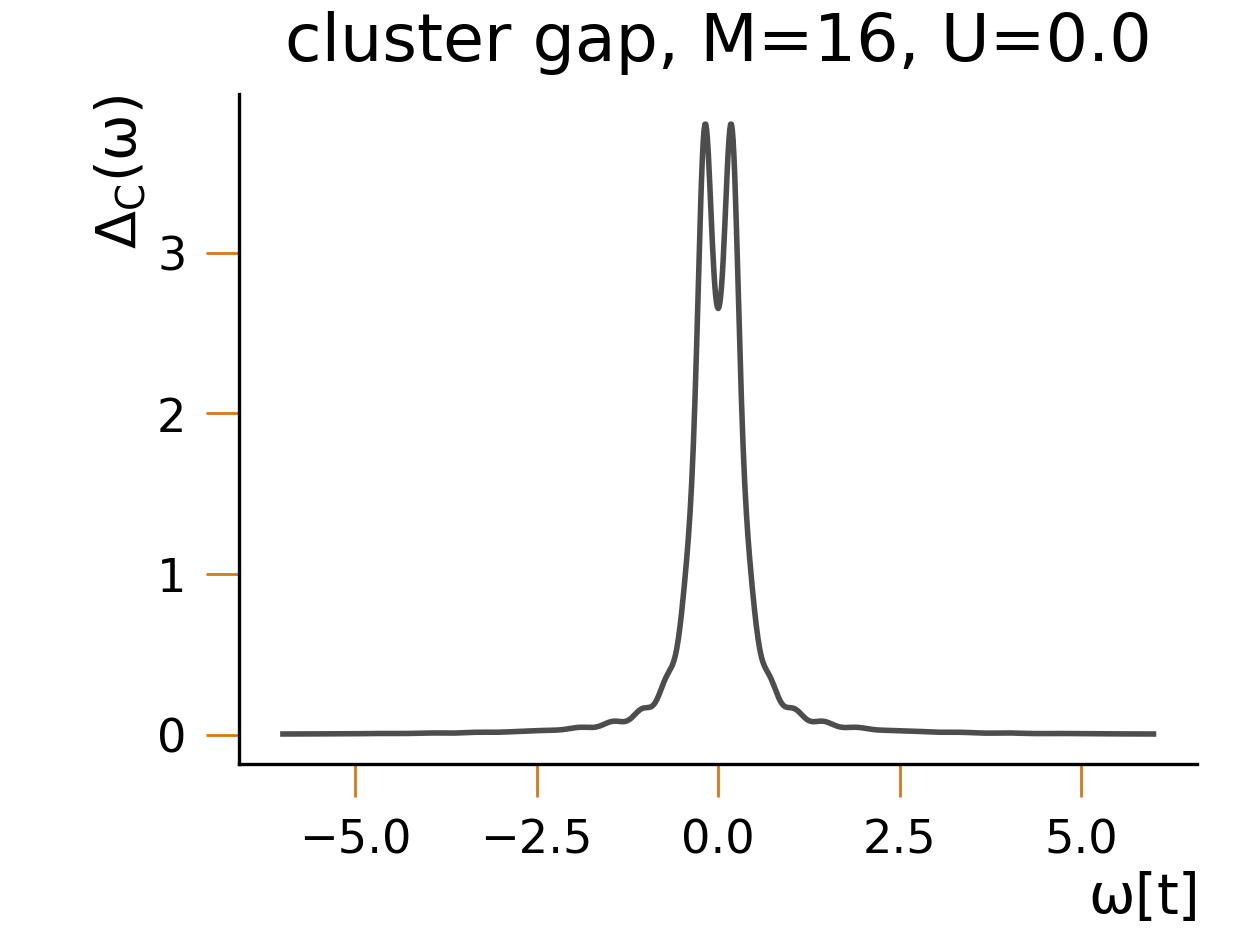}
    \caption{The average of the spectral function at the k points $k_{1/2} =
    \frac{M}{2} \pm 1$ for the M=16 site calculations as shown in Fig.~\ref{fig:clusterGF U0 small eta}.
    }
    \label{fig:averaged_spectral_function_U0}
\end{figure}
\begin{figure}[h!]
    \centering
    \includegraphics[width=0.45\textwidth]{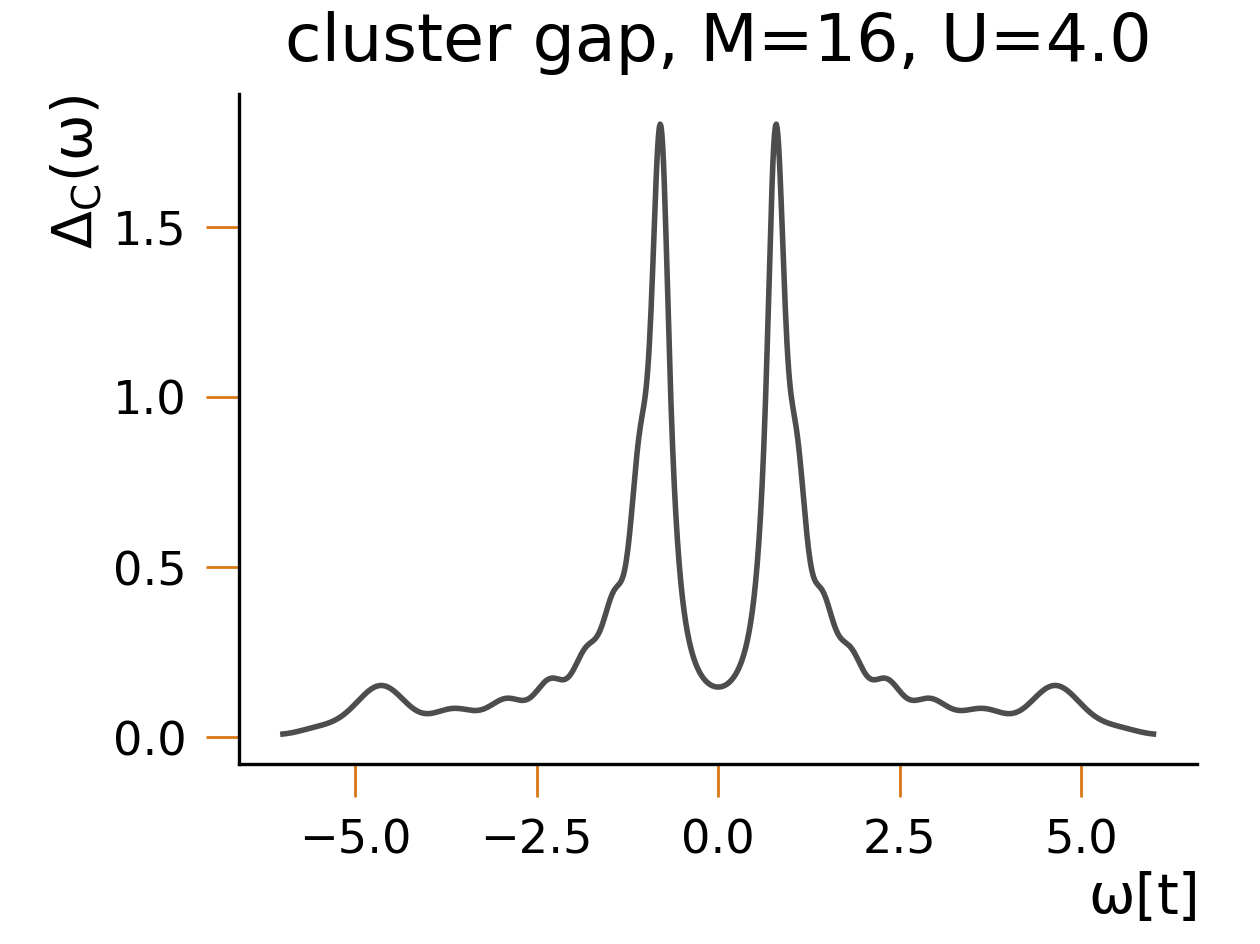}
    \caption{The same plot as in Fig.~\ref{fig:averaged_spectral_function_U0}
    but with an interaction strength of U=4t. We can see the shift in the peak
    positions which corresponds to the Hubbard gap.}
    \label{fig:averaged_spectral_function_U4}
\end{figure}
\begin{table}[t]
    \centering
    \begin{tabular}{|c|c||c | c | c | c | c|}
        \hline
        & M & U=1t & U=2t & U=4t & U=8t \\ [0.5ex]
        \hline
        6 & $(\Delta(U) - \Delta(0))/2$ & 0.03 & 0.16 & 0.66 & 2.24  \\
        \hline
        8 & $(\Delta(U) - \Delta(0))/2$ & 0.03 & 0.15 & 0.64 & 2.24 \\
        \hline
        10 & $(\Delta(U) - \Delta(0))/2$ & 0.02 & 0.14 & 0.63 & 2.24  \\
        \hline
        12 & $(\Delta(U) - \Delta(0))/2$ & 0.02 & 0.13 & 0.62 & 2.25  \\
        \hline
        14 & $(\Delta(U) - \Delta(0))/2$ & 0.02 & 0.13 & 0.62 & 2.26 \\
        \hline
        16 & $(\Delta(U) - \Delta(0))/2$ & 0.03 & 0.13 & 0.63 & 2.30  \\
        \hline
    \end{tabular}
    \caption{Half of the band gap $\Delta$, calculated using the peak positions from
    the interacting case and correcting them using the peak positions of the non
    interacting one.}
    \label{tab: Band gap 1D cluster eta corrected}
\end{table}
Now in order to arrive at the same result using the full CPT calculation, we
can simply plot the spectral function in the middle of the spectrum. If
we do this for multiple cluster sizes (Fig.~\ref{fig:cluster size first order U4}),
we can see, that the peaks slightly
move to the center with increasing cluster size. Hence we can extrapolate
the peak positions for the gap of an infinite cluster as shown in
Fig.~\ref{fig:finite size extrapolation U4}.
This results in gaps comparable to results obtained directly from the cluster (Tab.~\ref{tab: Band gap 1D cluster eta corrected}).
As in
the case where we just considered the cluster we see that the error is too large
to judge the gap accurately at $U=1t ,2t$.
\begin{table}[t]
    \centering
    \begin{tabular}{|c||c | c | c | c|}
        \hline
        & U=1t & U=2t & U=4t & U=8t \\ [0.5ex]
        \hline
        $\Delta_\text{Bethe}/2$ & 0.003 & 0.086 & 0.643 & 2.340 \\
        \hline
        $\Delta_{CPT}/2$ & 0.028 & 0.143 & 0.683 & 2.344 \\
        \hline
        $(\Delta_{CPT}/2)_{Error}$ & 0.025 & 0.057 & 0.040 & 0.004 \\
        \hline
    \end{tabular}
    \caption{In row one, we show half the band gap $\Delta$ calculated using Bethe ansatz. In
    row two, we give half the gap size obtained via CPT and in row three the extrapolation
    of the deviation. We can see that the results deviate on the order of magnitude of
    $10^{-1}$, as they already did, when we were just considering the cluster result. This was
    expected, since the accuracy of CPT is determined by the finite cluster.}
    \label{tab: Band gap 1D}
\end{table}
After this discussion, we now return to the 2D case which sparked this investigation in the first place.
We come to the conclusion that CPT is not able to resolve reliably the spectral weight within the gap and,
in particular, to predict the presence or absence of a pseudogap despite the resolution that the plot
in Fig.~\ref{fig:direct 4x4 U4 eta0.5 pseudogap band structure} suggests. This is, firstly, due to the
influence of the finite broadening parameter used to dampen the Gibbs oscillation induced by the Chebyshev
approximation, and, secondly, due to the finite level spacing associated with the chosen 2D cluster.

\subsection{Use Cases for CPT}
\label{sec:Applications}

\begin{figure}[h!]
    \centering
    \includegraphics[width=0.45\textwidth]{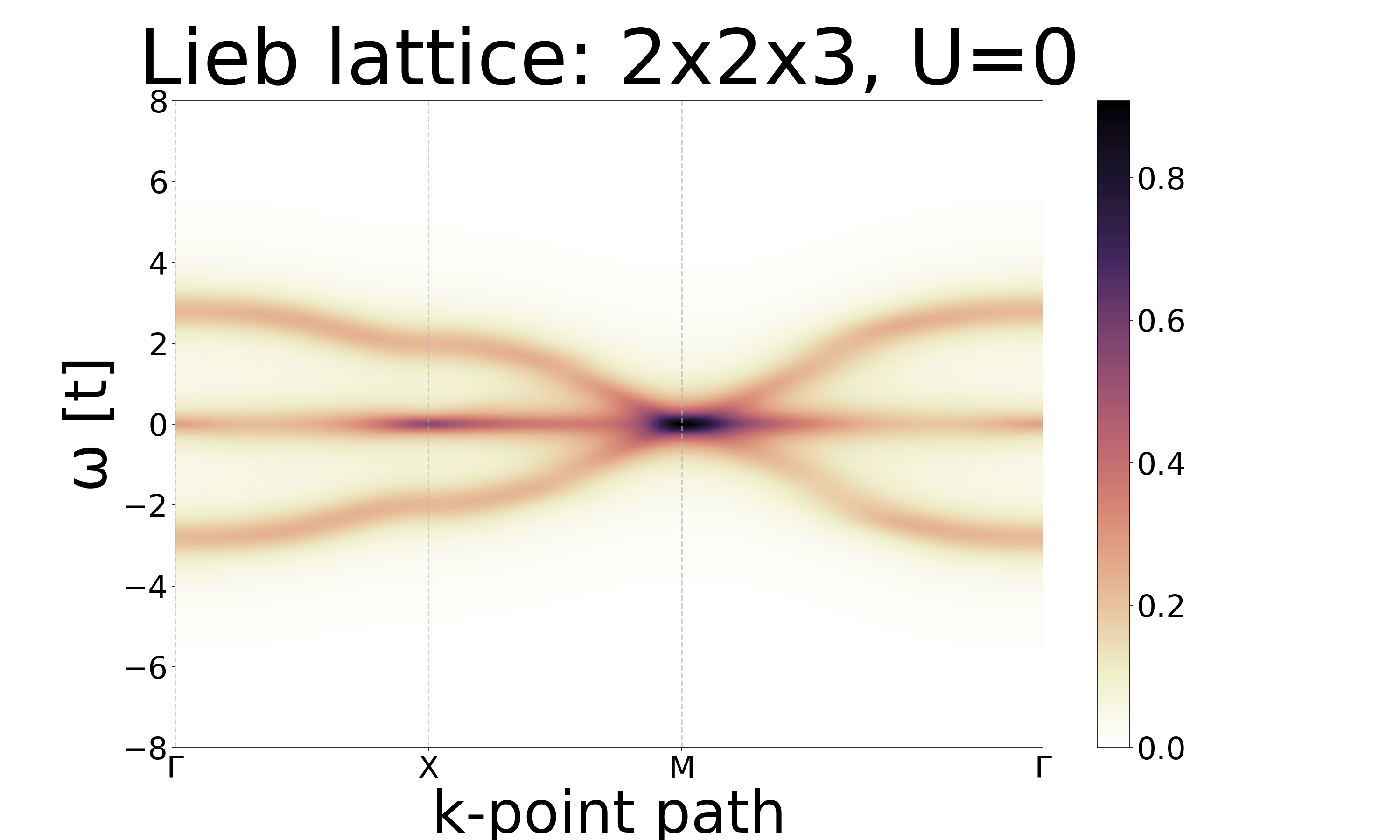}
    \caption{Spectral function of the Lieb lattice at $U=0$ obtained via CPT using
    a 2x2 cluster of three atomic unitcells. We can see the flat band at $\omega=0$.}
    \label{fig:liebU0}
\end{figure}

\begin{figure}[h!]
    \centering
    \includegraphics[width=0.45\textwidth]{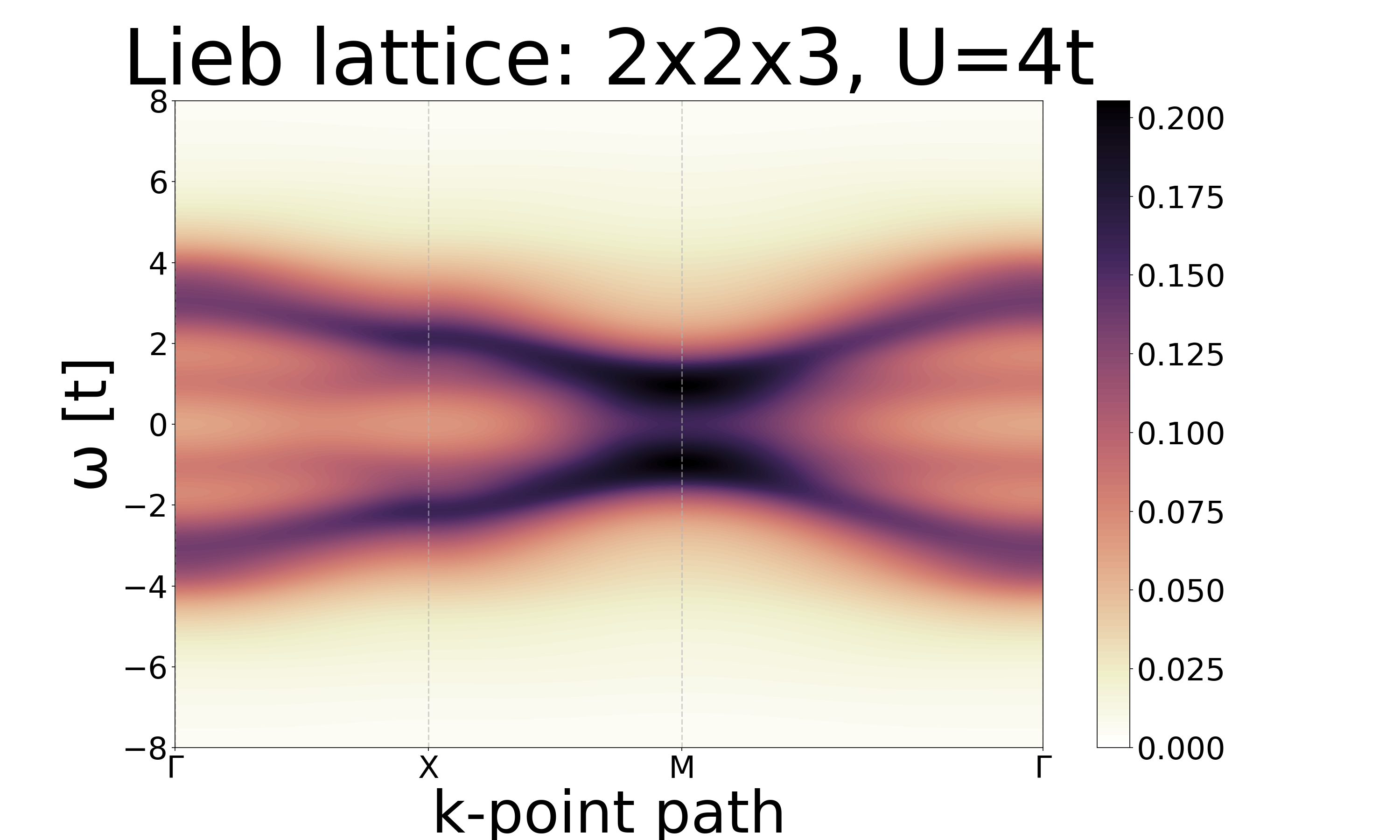}
    \caption{The CPT result for the spectral function of the Lieb lattice, but at
    $U=4t$. We can observe a splitting of the flat band.}
    \label{fig:liebU4}
\end{figure}

One might now pose the question as to why one should employ CPT in the
first place, since all of the reliable information is already contained in
the cluster results.
To this end, one should realize that just judging from the cluster
results it can be very hard to identify how the actual band structure of
a system would look like. CPT though, acting as a sort of interpolation
scheme between the single particle levels of the cluster and the infinitely large lattice, can
give a very good first guess of the band structure, which comes at a
negligible additional computational cost. Especially for materials, where
short range correlations are predominant, this guess
will be highly accurate.
Hence one might view it as a useful tool when scanning through a variety
of materials. One can simply perform fast CPT calculations for each material,
identify interesting band structures and then use more advanced methods
to investigate these materials further. As an example, we provide the spectral
functions for a more exotic lattice, the Lieb lattice \cite{weeks_topological_2010}
in Fig. \ref{fig:liebU0} and \ref{fig:liebU4}.

\section{Conclusion}
\label{sec:Conclusion}
In this study we have presented fundamental limitations of the CPT method.
Most importantly we showed that the cluster Green's function already contains, as far as interaction effects are concerned, all
the information that is included in the CPT Green's function and CPT only
stays consistent with this information. Analysing the 1D case we have shown that
the current computational limitations prohibit us to make accurate judgements of
features like the Hubbard gap at intermediate interaction strengths, as this would require
calculations on larger clusters, as the resolution is limited by the finite size
induced level splitting. Additionally and specifically, for the approach of calculating the cluster Green's
function via a Chebyshev expansion, we have shown that Gibbs oscillations require
the choice of a large broadening parameter that in return prohibits one from making
accurate judgements about the broadness of the peaks in the single particle spectrum.
Hence we conclude that use cases for CPT are limited to cases
where one is interested in obtaining numerically cheap initial guesses for
the spectral function of materials with short range correlations, while more advanced methods need to
be employed, to gain higher resolution.

\textbf{Data availability} The data presented in this
publication is avaliable on Zenodo under the DOI: 10.5281/zenodo.8063247.

\textbf{Funding} This project was supported by BMBF via the MANIQU grant no. 13N15576.

\bibliographystyle{unsrt}
\bibliography{cpt-bibliography}

\end{document}